\documentstyle[preprint,aps]{revtex}
\begin{document}
\draft \title{Polarization-Dependence of Born Effective Charge
              and Dielectric Constant in KNbO$_3$}
\author{Cheng-Zhang Wang, Rici Yu and Henry Krakauer}
\address{Department of Physics,
  College of William and Mary, Williamsburg, Virginia 23187-8795}
\date{Received 11 March 1996; revised manuscript received 26 June 1996 }
 \maketitle
\begin{abstract}
  The Born effective charge Z$^{*}$ and dielectric tensor
  $\epsilon_{\infty}$ of KNbO$_3$ are found to be very sensitive
  to the atomic geometry, changing by as much as 27$\%$ between
  the paraelectric cubic and ferroelectric tetragonal and rhombohedral
  phases. Subtracting the bare ionic contribution reveals changes
  of the dynamic component of Z$^{*}$ as large as 50$\%$, for atomic
  displacements that are typically only a few percent of the lattice
  constant. Z$^{*}$, $\epsilon_{\infty}$ and all phonon frequencies
  at the Brillouin zone center were calculated using
  the {\it ab initio} linearized augmented plane-wave 
  linear response method with respect to the
  reference cubic,  experimental tetragonal,
  and theoretically determined rhombohedral
  ground state structures.  The ground state rhombohedral structure of
  KNbO$_3$ was determined by minimizing the forces on the relaxed atoms.
  By contrast with the cubic structure, all
  zone center phonon modes of the rhombohedral structure
  are stable and their frequencies are in
  good agreement with experiment. In the tetragonal phase,
  one of the soft zone center modes in the cubic phase
  is stablized.
  In view of the small atomic
  displacements involved in the ferroelectric
  transitions, it is evident that not
  only the soft mode frequencies but also the Born effective charge
  and dielectric constants are very sensitive to the atomic
  geometry.

\end{abstract}
\pacs{ PACS number: 77.80.e  77.22.Ej  63.20.Dj }
\newpage \narrowtext

\section{Introduction}

Potassium niobate (KNbO$_3$) is one of the most studied members in the
important class of perovskite structure ferroelectrics.
Like BaTiO$_3$,  KNbO$_3$
undergoes three successive ferroelectric phase transitions as the
temperature is decreased.\cite{fontana84}
It transforms from  the cubic to a
tetragonal phase at 701 K, to an orthorhombic phase  at 488 K,
and finally to a rhombohedral phase at 210 K.
Despite decades of studies on these perovskite systems,
there is  still controversy as to the structure of the
high temperature phases and the related character of
the phase transitions, with contradictory evidence for
order-disorder \cite{comes68-dougherty92} versus
soft-mode\cite{nunes71-samara87} behavior.
The  phase transitions and macroscopic polarization are strongly sensitive to
chemical compositions, defects, details  of domain structure,
and stresses, which  complicates both theoretical and
experimental studies. Nevertheless, considerable insight has been
gained in the past few years into the microscopic mechanism for
the ferroelectric instability
from first principles calculations, based on the local
density functional theory.
\cite{cohen-k-90,cohen92,singh-b-92,king-s94,zhong95,yu95}
Cohen and Krakauer\cite{cohen-k-90} performed a series of total
energy calculations on BaTiO$_{3}$ and showed that
the cubic phase is unstable against zone center distortions.
They also  showed that hybridization between oxygen $2p$ and transition
metal titanium $3d$ electrons is important feature in explaining  the
ferroelectric instabilities in BaTiO$_{3}$ and PbTiO$_{3}$.\cite{cohen92}
King-Smith and Vanderbilt\cite{king-s94} carried out systematic total energy
calculations for zone center distortions for eight perovskite oxides.
Using an effective Hamiltonian constructed from first principles calculations,
Monte Carlo simulations for BaTiO$_3$
by Zhong, Vanderbilt and Rabe\cite{zhong95}
obtained the phase sequence, transition temperatures, latent heats,
and spontaneous polarizations, which are in good agreement with
experiments.  With the use of a linear response density functional method,
Yu, Krakauer, and Wang \cite{yu95} obtained a complete mapping in the Brillouin
zone of the structural instability for cubic KNbO$_{3}$, revealing
a pronounced two dimensional character, corresponding,
in real space, to
chains along the \mbox{[1 0 0]} directions of atoms
coherently displaced along the chain direction.

In this paper, we determine
the zero-temperature ground state rhombohedral structure of KNbO$_{3}$
by minimizing the forces acting on the relaxed atoms.
Results are then presented of linear response calculations of
zone center phonon spectra,
Born effective charges, and dielectric constants for
the cubic,  tetragonal, and theoretical
ground state rhombohedral structures, demonstrating that
not only the soft mode frequencies but also the Born effective
charges and dielectric constants are very sensitive to the atomic geometry.

\section{Methods}

Total energy and force  calculations were performed within the local
density approximation (LDA), using the  LAPW (linearized-augmented-plane-wave)
method.\cite{lapw75-2} The ground state structure
was determined by minimizing the forces\cite{yu91} on the relaxed atoms.
Phonon spectra,  Born effective charge and dielectric tensors were
obtained with the recently developed LAPW linear response
method\cite{yu95,yu94,wang94}.
We refer the reader to Ref.\cite{yu94} for details regarding this  method.
The dynamical matrix at a wavevector
${\bf q}$ is calculated from the first order changes in the wavefunctions due
to a
phonon-like
perturbation, $w_{i\alpha}({\bf R}) = w_{i\alpha} e^{i {\bf q} \cdot {\bf R}}$,
where $w_{i\alpha}({\bf R})$ indicates the Cartesian $\alpha$ component
of the displacement of atom $i$ in the unit cell specified by the direct
lattice vector
${\bf R}$;
the phonon frequencies are then obtained by standard matrix diagonalization.
The Born effective charge and dielectric tensors are obtained in the
long wavelength limit as
\begin{equation}
  \hat{{\bf q}} \cdot \epsilon_{\infty} \cdot \hat{{\bf q}}
  = \lim _{{ \bf q \rightarrow 0} } \biggl [ 1- \frac{V_{ind}({\bf q })}
  {V_{total}({\bf q }) } \biggl ],
\end{equation}
\begin{equation}
  Z^{*}_{\alpha, \beta}(i)= Z_{i} + \Omega \frac{ \partial P_{\alpha} }
  {\partial \omega_{i\beta} }
  \biggl |_{ {\bf E =0} } ,    \label{z-b-eff}
\end{equation}
where $\hat{{\bf q}}$ is the unit vector in the direction of
wavevector, $V_{ind}$ is the induced potential when  the macroscopic field is
applied,
$Z_{i}$ is the bare ionic charge of the $i$th ion, $ \Omega $ is the volume of
the unit cell, and $P_{\alpha}$ is the Cartesian $\alpha$ component of
the macroscopic electronic  polarization under the phonon-like perturbation
while
constraining the macroscopic electric  field to be zero.
We calculated the Born effective charge and dielectric tensors
at very small values of $q$ $\sim$ 0.01 2$\pi/a$, where $ a$ is the lattice
constant.
Kerker type pseudopotentials\cite{kerker}
were used to bypass  the need to treat  the chemically inert
localized inner core states.
The  shallow  semicore states K(3$s$) and Nb(4$s$,4$p$) were
pseudized  and included in a lower energy window of a two-window  variational
calculation.
The Wigner interpolation formula\cite{wigner} was used
for the exchange-correlation potential.
A uniform (6 6 6) $k$-point set\cite{monk} was
used  in most calculations to sample the Brillouin zone (BZ) integration,
yielding 10 special $k$-points in the irreducible BZ wedge for the cubic
structure, 18 special $k$-points for the tetragonal structure,
 and 28 special $k$-points for the  rhombohedral structure.
The muffin-tin radii were chosen as 2.50 a.u.(atomic  unit) for K, 1.92 a.u.
for Nb
and 1.58 a.u. for O.  The kinetic energy cutoff was  22.5 Ry, yielding
approximately 810 LAPW basis functions at each $k$-point.

For cubic KNbO$_3$, we obtained the theoretical equilibrium lattice constant,
$a_0$=4.00\AA, with bulk modulus, $B_0$=194 GPa,
and pressure derivative, $B^{'}_{0}$=4.3,
by fitting the LAPW total energy calculations to the
Birch-Murnaghan equation of state.\cite{birch} This lattice constant is
only 0.4$\%$ smaller than the  experimental value 4.016 \AA.\cite{fontana84}
The change in total energy due to ferroelectric distortions
is very sensitive to volume\cite{cohen-k-90,singh-b-92},
so despite this small difference the
experimental lattice constant  was used in all the calculations described
below.
Our lattice constant is
slightly larger than  previous LDA
calculations\cite{king-s94,singh95-knbo3}.
 This  difference is probably due to  the use of different forms
for exchange-correlation potential. As also seen in other
systems\cite{wang95-1,lu89}, the Wigner form tends to yield slightly
larger equilibrium lattice constants than other forms, and as expected
the bulk modulus is accordingly slightly smaller.

\section{Ground State Structure}

To determine the lowest energy  geometry,
the atomic positions were  relaxed, minimizing the forces on all atoms while
constraining the system to have at least rhombohedral symmetry.
This was done using a cubic unit cell,
based on the experimental observation that the rhombohedral strains
are extremely small (the sheer strain angle is merely 11$^{'}$\cite{hewat73}).
We checked that distorting the cubic cell with the experimentally
observed rhombohedral strain results in only slightly larger residual forces on
the atoms, indicating that the strain has little effect on the
relative atomic displacements.

In units of the lattice constant (4.016\AA), the atomic positions
in the reference cubic perovskite structure
were taken as \mbox{(0.5, 0.5, 0.5)} for K,
\mbox{(0, 0, 0)} for Nb
and \mbox{(0.5, 0, 0)}, \mbox{(0, 0.5, 0)}, and \mbox{(0, 0, 0.5)}
 for the three O atoms O$_{1}$,
O$_{2}$ and O$_{3}$,
respectively.
Without any loss of generality, the Nb atom was chosen
as a fixed reference point, with respect to which
the K atom was displaced to \mbox{(0.5+$\delta_{K}$, 0.5+$\delta_{K}$,
0.5+$\delta_{K}$)} and the oxygen atoms to
\mbox{(0.5$ +\delta_{I}$, $\delta_{II}$, $\delta_{II}$)}
and symmetric permutations, where  $\delta_{K}$, $\delta_{I}$ and $\delta_{II}$
are three
independent parameters.
The process of structural optimization was terminated when the maximum
residual force was less than $\sim 0.3$ mRy/a.u., i.e. 8 meV/\AA.

Table~\ref{ground-s} presents our results of ground state structures
with  three different $k$-point sets.
 The energy difference between the ground and cubic
states are also shown. Previous LDA calculations
with LMTO (linearized-muffin-tin-orbitals) method\cite{postni93}
and experimental measurements are also included for
comparison. Table \ref{ground-s} reveals that the structural
parameters are essentially converged using a (6 6 6)  $k$-point set.
Use of the \mbox{(6 6 6)} set increases
the difference between the rhombohedral
ground state and the ideal cubic structure energy, \mbox{ ($E_{g}-E_{c}$)},
by \mbox{200$\%$} compared with \mbox{(4 4 4)} set.
Increasing the BZ sampling to an \mbox{ (8 8 8)} set changes the energy
difference  by just 7$\%$ compared with the \mbox{(6 6 6)} set,
 suggesting that \mbox{(6 6 6)} set would be adequate in
many calculations.  This is confirmed by the fact
that the ground state structural parameters determined with \mbox{(6 6 6)}
and \mbox{(8 8 8)} sets exhibit little difference.
Negative values of \mbox{ ($E_{g}-E_{c}$)} mean that the
cubic structure is unstable against zone center distortions.

It should be emphasized that the ground state structure is not simply
a displacement of the Nb atoms against the rigid sublattices
of K and O. The relaxation of these sublattices significantly affects
the stability of the rhombohedral  structure.
For example, with the use of \mbox{(8 8 8)}
$k$-point set, moving the Nb atom from the origin to
\mbox{ (0.025, 0.025, 0.025)} (in units of the lattice constant),
while keeping other atoms fixed,
causes the total energy of the system  to decrease, but only  by 0.45 mRy.
Relaxing the K and O atoms sublattices causes the energy to decreases
by 1.99 mRy. Thus the displacement of the Nb sublattice against
the rigid lattice of all the  other atoms is not the lowest energy
configuration.
This conclusion contradicts the results of Postnikov {\it et
al.}\cite{postni93}.
As shown in Table~\ref{ground-s},
Postnikov {\it et al.} determined the ground state structure of
KNbO$_3$  at two  lattice parameters,
3.93\AA ~and 4.00\AA ~, using  the LMTO  method.\cite{postni93}
In both cases, they find K atom displacements that are  essentially the same
as rigid oxygen octahedron displacements ($\delta_K \approx
\delta_I=\delta_{II}$),
which means that the ground state structure
corresponds to an essentially pure distortion of Nb sublattice against
the rigid lattice of all the  other atoms. This  result  is strikingly
different
from our calculations.
This discrepancy cannot be fully
ascribed to the approximation of rigid oxygen octahedra used in their
calculations, since we find that the distortion of the oxygen octahedra
(i.e. the difference between  $\delta_I$ and $\delta_{II}$) is small.
Furthermore, their calculated zone center phonon eigenvectors for the cubic
structure with the same method
are quite different from the results of our calculations
and other's (see discussion in V).
Table~\ref{ground-s} also lists experimental values,
with the measured displacements $\delta_K$,
$\delta_I$ and $\delta_{II}$\cite{hewat73}
all about 20 $\%$ larger than our calculations.
This level of agreement is  satisfactory,
given the sensitive dependence of atomic positions
on the volume.

\section{Calculations for Z$^{*}$  and $\epsilon_{\infty}$}

The Born effective charge tensor Z$^{*}$
characterizes the influence of  long-range Coulomb interactions
on the vibrational and optical properties of ionic insulators.
First principles calculations
of these charges in perovskite ferroelectrics have become
available only very recently.\cite{yu95,resta93,zhong-z-94,ghosez95}
Here we report on calculations of Z$^{*}$ and $\epsilon_{\infty}$, the
electronic component of the static dielectric constant, for
the theoretical ground state rhombohedral structure and experimental tetragonal
structure\cite{tetra-exp}  of KNbO$_3$. Cubic phase  Z$^{*}$ were
previously reported.\cite{yu95}  We find that
the eigenvalues of these tensors
are quite different from their counterparts in the cubic structure.
To ensure the reliability of the comparison, we recalculated these quantities
for the  cubic structure with the  denser $k$-point sampling in BZ, i.e., the
\mbox{(6 6 6)} set that
is used for rhombohedral and tetragonal structures. We have examined the
convergence of the
Born effective  charges with respect to $k$-point sampling.
The \mbox{(6 6 6)} set
reduces the violation of the acoustic sum rule by an order of magnitude
compared with the \mbox{ (4 4 4)} set\cite{yu95}. Specifically, with
the use of the \mbox{(6 6 6)} $k$-point set, $\sum_{i}$Z$^{*}(i)$ is reduced
from
0.35 using the  (4 4 4) set to
0.03 for the cubic  structure; for the rhombohedral structure,
the diagonal and off-diagonal terms are reduced from 0.11 and 0.56
 to -0.07  and  0.02, respectively. For the tetragonal structure,
the three diagonal terms of  $\sum_{i}$Z$^{*}(i)$ are -0.02, -0.02 and 0.07
with the use of the (6 6 6) $k$-point set.
This level of convergence is comparable to that obtained in simple
semiconductors.\cite{wang95-1}
The errors due to the use of a small but finite wavevector
were  checked and found to be negligible.
For example, in the calculations of
Z$^{*}$ in the experimental tetragonal
structure, the use of a smaller wavevector \mbox{{\bf q}=(0, 0, 0.002)2$\pi/c$
}
yields the $\hat{z}$-component of Z$^{*}$(O$_3$)  to be -5.350, very close to
-5.348,
the value obtained with \mbox{ {\bf q}=(0, 0, 0.01)2$\pi/c$}.

In the  cubic structure, the Born effective charge tensors Z$^{*}$(K)
and Z$^{*}$(Nb) are both isotropic due to the high site symmetry.
At the O sites, however, there exist two inequivalent directions:
one along the \mbox{Nb-O} bond and the other perpendicular
to this bond, denoted by  $_{\parallel}$ and $_{\perp}$, respectively.
Thus, Z$^{*}$(O) is diagonal but with two
distinct values, Z$^{*}$(O)$_{\parallel}$ and Z$^{*}$(O)$_{\perp}$.
In the tetragonal structure, all Born effective charge tensors are diagonal,
and Z$^{*}$ of K, Nb and O$_{3}$ each have two distinct
values, whereas Z$^{*}$ of the equivalent O$_{1}$  and O$_{2}$ atoms
have three distinct values.
In the  rhombohedral phase, the Born effective charge tensors no longer have
a simple diagonal form due to   the lower  symmetry and we have found
\begin{equation}
  Z^{*}(K) = \left ( \begin{array}{rrr}
      1.14  &  -0.01   & -0.01  \\
      -0.01 &   1.14   &  -0.01  \\
      -0.01 & -0.01    & 1.14
    \end{array}       \right),
  \nonumber
\end{equation}
\begin{equation}
  Z^{*}(Nb) = \left ( \begin{array}{rrr}
      8.16   & -0.35  & -0.35  \\
      -0.35  & 8.16   & -0.35   \\
      -0.35  & -0.35  & 8.16
    \end{array}       \right),
  \nonumber
\end{equation}
and
\begin{equation}
  Z^{*}(O_{1}) = \left ( \begin{array}{rrr}
      -6.27 & 0.14  & 0.14 \\
      0.24 & -1.55  &  0.00 \\
      0.24 &  0.00  & -1.55
    \end{array}      \right).
  \nonumber
\end{equation}
For purpose of comparison, it is convenient to focus on the eigenvalues
of these tensors, presented in Table~\ref{z-compare}.
The Berry phase calculations for the cubic structure\cite{zhong-z-94}
and the experimental tetragonal structure\cite{resta93,zhong-z-94} are also
included for comparison.

There is no requirement that the Born effective
charge tensor be symmetric, because it is a mixed second derivative of the
total energy,
$\cal E$$_{total}$, with respect to
macroscopic electric field component, E$_{\alpha}$, and atomic
displacement component, $\tau_{i,\beta}$:
Z$^*(i)$ = $\partial^{2} {\cal E}_{total} / \partial E_{\alpha}\partial
\tau_{i,\beta} $.
Thus the Born effective charge
can either be regarded as the derivative of the polarization with
respect to atomic displacement at zero macroscopic field, or as the
derivative of the force on an atom with respect to the macroscopic
field at zero atomic displacement.
The Born effective charge tensors for  K and Nb in  the
rhombohedral structure are symmetric and the macroscopic polarization
direction,  \mbox{[1 1 1]},
is an obvious principal axis, with corresponding
eigenvalues of Z$^{*}$(K)=1.13 and Z$^{*}$(Nb)=7.47 respectively.  The other
two perpendicular eigenvectors with degenerate eigenvalues
may be chosen as \mbox{[1 $\bar{1}$ 0]} and \mbox{[1 1 $\bar{2}$]}.
For the three equivalent oxygen atoms, the three eigenvectors of the Born
effective charge tensors are no longer perpendicular to each other
since the tensors are no longer symmetric.
For example,  for the  O$_{1}$ atom at
(0.5$+\delta_{I}$, $\delta_{II}$, $\delta_{II}$) in the rhombohedral structure,
\mbox{[0 1 $\bar{1}$]} is an eigenvector direction  and the
corresponding  eigenvalue is Z$^{*}$(O$_1$)= -1.55 as shown in
Table~\ref{z-compare},
but the other two eigenvectors cannot be determined by
symmetry alone and depend on the details of interactions in the
material.  Nevertheless, we still
use Z$^{*}$(O)$_{\parallel}$ and Z$^{*}$(O)$_{\perp}$ to denote the
eigenvalues for convenience, although the corresponding eigenvectors
deviate slightly  from their counterparts in the cubic structure.

All of the LDA calculations in Table~\ref{z-compare}
 show unusually large Born effective charges
Z$^{*}$(Nb) and Z$^{*}$(O)$_{\parallel}$.  These values are by far
larger than their corresponding nominal ionic charges $+5$ and $-2$,
revealing large dynamic charge transfers
along the \mbox{Nb-O} bond as the length of the bond is varied.
These large Born effective charges result from the strong
covalent interactions between transition metal  and the
oxygen atoms in these materials.\cite{cohen-k-90}
This has been demonstrated  recently by
Posternak, Resta, and Baldereschi,\cite{posternak94}
who observed that the unusually large Born effective charges are
reduced to their nominal values when the covalence between Nb 4$d$ and
O 2$p$ orbitals  is artificially suppressed.
These large Born effective charges
give rise to strong long-range ionic interactions, which favor the
ferroelectric instability.\cite{cohen-k-90} It is not surprising, therefore,
that ferroelectricity is  sensitive to the size and
formation of domains as well as electric boundary conditions.

Our calculations show that the eigenvalues of the Born effective charge
tensors,
especially Z$^{*}$(O)$_{\parallel}$ and the component of Z$^{*}$(Nb)
along macroscopic polarization direction, are quite
sensitive to the changes in atomic geometry.  The atomic displacements involved
in the transformation from the cubic to the tetragonal and ground state
rhombohedral structures
are rather small, typically only a few percent of the lattice constant.
Nevertheless, the values of Z$^{*}$(Nb) and Z$^{*}$(O)$_{\parallel}$ are
reduced  by up to
23$\%$ in the rhombohedral structure and by about  27$\%$ in the tetragonal
structure. Subtracting the bare ionic charges associated with
the atoms (+5 for Nb and -2 for O)
shows that the change of the dynamic charge component of Z$^{*}$
can be as large as 50$\%$. This sensitivity  originates from the
strong dependence\cite{posternak94}
of the Nb 4$d$ and O 2$p$  hybridization on the ferroelectric
distortion, which changes the Nb-O bond length and site symmetry of atoms
in crystal.
As shown in Table \ref{z-compare}, Zhong $et$ $al.$'s 
Berry phase calculations agree with our results within
a few percent for the cubic structure.
It should be noted that in their calculation for the tetragonal
phase\cite{zhong-z-94}, they used an ideal tetragonal structure with
experimental lattice constants but without observed internal strains.
Resta {\it et al.}\cite{resta93} extracted Born effective charges for the
tetragonal phase from the finite differences
of polarization and argued that
the Born effective charges are approximately independent of the  atomic
displacements
and that the macroscopic polarization is therefore linear with the internal
strain.
Our results differ from this.
We find  that the Born effective charges strongly depend on the structural
details.
Recently Ghosez {\it et al.}\cite{ghosez95}
calculated the  Born effective charge
tensors for different phases of BaTiO$_{3}$, and found that
they are also strongly dependent  on the atomic positions, in agreement with
the present calculations for KNbO$_3$.

In the absence of macroscopic strain, symmetry requires that the
changes in the Born effective charge are even functions of the
amplitude of the internal displacements, when expanded about the cubic
structure. In the absence of internal displacements, the volume
dependence and macroscopic strain, which are important in minimizing
the total energy, are found to have little influence on the Born
effective charges. Reducing the lattice constant by 0.47$\%$ in the
cubic structure (from 4.016 {\AA} to 3.997 {\AA}), Z$^{*}$(Nb)
decreases by only 0.85$\%$. The effect of a pure tetragonal strain is
examined in a calculation in which $a=b=$ 3.997 {\AA} and all atoms
are kept at their ideal positions, {\it i.e.}  there are no internal
strains. For $c/a$ ratios of 0.9835, 1.0, and 1.0165 (the last being
the strain in the observed tetragonal structure), the values of the
$z$-component of Z$^{*}$(Nb) are 9.626, 9.586, and 9.556,
respectively, with the overall change being only
0.7$\%$. Thus, the variation of the Born effective charges between
different atomic geometries reported above, are almost entirely due to
the internal strains.

The polarization change is defined as
\begin{equation}
\Delta {\bf P} = {\bf P}( \tau) -  {\bf P}( 0) =
      \sum_{i} \int_{ 0}^{ \tau} \frac{e}{\Omega} Z^{*}(i, u)du,
  \label{polarization-change}
\end{equation}
where  $u$ is a symbolic parameter characterizing
the structure and 0 and $\tau$
correspond to  the  starting and ending structures;
$i$ is the atomic index and $\Omega$ is the volume of unit cell.
In order to estimate the integral in Eq. (\ref{polarization-change}),
we consider two approximations to estimate the range of macroscopic
polarization for the experimental tetragonal structure. This is done
by  ignoring the $u$-dependence of Born effective charge tensors
and instead using our calculated values of
Z$^{*}$  for (i) the cubic structure,  and (ii) the tetragonal structure,
yielding
two different values of $\Delta {\bf P}$,  0.44 C/m$^2$ and 0.33 C/m$^2$,
respectively.
The measured value is 0.37  C/m$^2$,\cite{kleemann84} close to
the average of the two estimates above.
The large difference between the
two estimated values  demonstrates that the ferroelectricity
is essentially a nonlinear phenomenon.

We have calculated the dielectric tensors $\epsilon_{\infty}$ for the
cubic, experimental tetragonal, and theoretical ground state
rhombohedral structures and  presented the
eigenvalues  in Table~\ref{tensor-eps-eig-compare}.
The cubic structure has an isotropic dielectric tensor,
$\epsilon_{\infty}$ = 6.63.
The tetragonal structure has a diagonal dielectric tensor,
and the eigenvalue in the direction of polarization [ 0 0 1] is 5.07,
about  a 24\% reduction from the
cubic phase. For the rhombohedral structure,
the dielectric tensor has a symmetric form
and $\epsilon_{\infty,11}
=\epsilon_{\infty,22}=\epsilon_{\infty,33}=5.79$ and
 $\epsilon_{\infty,12}
= \epsilon_{\infty,13}= \epsilon_{\infty,23}=-0.15$.
The three eigenvalues are 5.93, 5.93, and 5.49,
with eigenvectors along [1 $\bar{1}$ 0], [1 1 $\bar{2}$] and [1 1 1],
respectively.
Compared to the cubic phase, the largest reduction of eigenvalues
is 17$\%$, also along the direction of polarization. Both the tetragonal and
rhombohedral cases show that
the dielectric tensors also strongly depend on  atomic  geometry.

\section{Zone center phonon calculations}

Due to the lower symmetry of the reference rhombohedral and tetragonal
structures, the linear response calculations take significantly longer
compared to similar calculations using the cubic phase as the reference
structure.
For each phonon wavevector {\bf  q}, the first order change in the
wavefunctions must
be calculated at many more k-points in the Brillouin zone due to the
lower symmetry. For this reason,
we have limited our calculations of the dynamical matrix in the rhombohedral
and tetragonal phases to the zone center, {\bf q}={\bf 0}.
Table~\ref{knbo3-c-lto} compares our zone center phonon frequencies
of the cubic structure with the frozen phonon
LAPW calculations of Singh and Boyer\cite{singh-b-92},
planewave pseudopotential calculations of
Zhong, King-Smith, and Vanderbilt\cite{zhong-z-94},
and LMTO calculations of Postnikov {\it et al.}\cite{postni94-1}
In the cubic phase the twelve optical phonons at {\bf q}={\bf 0} are
classified as three F$_{1u}$ and one F$_{2u}$ modes, each of which is
triply degenerate. The F$_{2u}$ mode is labelled 4 whereas the other three
F$_{1u}$ modes are labelled as 1, 2, and 3 in order of increasing frequency.
All the calculations find unstable TO modes at the zone center with
similar imaginary frequencies corresponding to the observed soft mode.
The longitudinal optic (LO) phonon frequencies were obtained from a
dynamical matrix which is a combination of
a zone center dynamical matrix without macroscopic field, $D^{TO}$, and
a term arising from the long range Coulomb interaction:
\begin{equation}
  D^{LO}_{i\alpha,j\beta} =D^{TO}_{i\alpha,j\beta}
        +    \frac { 4\pi e^{2}  }
  {\Omega  \sqrt{M_i M_j} }
  \frac{ ({\bf q \cdot Z^{*}}(i))_{\alpha}
    ({\bf q \cdot Z^{*}}(j))_{\beta}    }
  { {\bf q}\cdot \epsilon_{\infty} \cdot {\bf q} }
  \label{dm-lo-to} ,
\end{equation}
where $M_i$ and  {\bf Z}$^{*}(i)$ are  the mass and Born effective charge
tensor of atom $i$,
${\bf \epsilon}_{\infty}$ is the dielectric tensor,
$\Omega$ is the volume of unit cell, $\alpha,\beta$ are Cartesian indices.
This partitioning of the dynamical matrix is necessary
precisely at the zone center, {\bf q}={\bf 0}, because the distinction between
transverse and longitudinal modes becomes delicate.
At finite wavevectors, we can directly calculate the full dynamical matrix from
the first order forces on the atoms.
Frequencies calculated  this way for very small wavevectors q $\sim$ 0.01
2$\pi/a$
agree with Eq.~(\ref{dm-lo-to}), showing the internal consistency of our linear
response calculations.
All LO modes are stable due to the contribution of the macroscopic field.
The TO$_4$ mode with frequency 243 cm$^{-1}$ is both Raman and infrared
inactive,
and thus does not
exhibit LO-TO splitting.  Zhong, King-Smith, and Vanderbilt obtained
their LO frequencies by using $\epsilon_{\infty}$=4.69, extracted from
experiment,
whereas we used both $\epsilon_{\infty}$=4.69 and
our larger calculated dielectric constant $\epsilon_{\infty}$=6.63. The highest
LO$_3$
frequency increases by 131 cm$^{-1}$ when  the  smaller dielectric constant
$\epsilon_{\infty}$=4.69 is used.
Compared to our previous results with (4 4 4) $k$-point set\cite{yu95}, the
soft mode frequencies is found to be more unstable by about 50$i$ cm$^{-1}$,
consistent with the observation in Table I that (6 6 6) $k$-point  set
yields a deeper well depth. The other modes are relatively unaffected.

We present the eigenvectors of zone center optic phonon modes
 in the cubic structure in Table~\ref{knbo3-c-to-vector}.
Note that the eigenvectors are the actual displacements
 weighted by the square root of atomic mass.
The basic features of the TO phonon eigenvectors are as follows.
In the soft TO$_{1}$ mode, Nb vibrates against three oxygen atoms with K
essentially unmoved. In  the TO$_{2}$ mode, K vibrates against all of the
other atoms. The TO$_{3}$ mode is dominated by the vibration of one O atom
against the  other two  O atoms, whereas  Nb and K atoms
oscillate with only slight amplitudes.   In the  TO$_{4}$ mode, K and
Nb are exactly at rest,  and the oscillations involve only O atoms.
When the TO$_{4}$ mode is viewed along the $\hat{z}$ direction
as shown in Table~\ref{knbo3-c-to-vector}, the O$_1$
at (0.5$a$, 0, 0) and O$_2$ at (0, 0.5$a$, 0)
move with the same amplitudes but in opposite directions, and
O$_3$ at (0, 0, 0.5$a$) is undisplaced.
This mode has no instantaneous dipole moment so the TO$_4$ phonon is
infrared inactive.
LO and TO modes have quite different eigenvectors, as previously noted by
Zhong {\it et al}\cite{zhong-z-94}.  The correlation between
the infrared active modes  may be characterized by the eigenvector
overlap matrix
A$_{ij}$ = $<u_i^{TO}|u_j^{LO}>$ and we have
\begin{equation}
  A = \left ( \begin{array}{ccc} 0.63 & 0.13 & 0.77 \\
      0.10 & 0.99 & 0.10 \\
      0.78 & 0.02 & 0.62
    \end{array}       \right).
  \nonumber
\end{equation}
The matrix A indicates that there  exists no one-to-one correspondence
between the TO and LO modes in the perovskite ferroelectrics and that
the soft TO$_1$ mode is most closely
coupled with the highest frequency mode LO$_3$.\cite{zhong-z-94}

Our zone center optical  phonon modes  agree well with the
LAPW frozen phonon calculations by Singh and
Boyer\cite{singh-b-92}, but differ significantly
from those by Postnikov {\it et al.}\cite{postni94-1}
using the LMTO method. One of the striking differences lies in the
eigenvectors of the soft TO$_1$ modes.
We observed that  K is essentially unmoved in the soft modes, while
Postnikov {\it et al.}\cite{postni94-1} observed that K atom tends to
move with the oxygen octahedron.
Table \ref{soft-mode} displays the relative atomic displacements
in the soft mode from LDA calculations, as well as the
experimental values from the observed tetragonal structure.
Both frozen phonon calculations of the soft TO$_1$ mode with
LAPW method\cite{singh-b-92} and  plane wave method\cite{zhong-private} yield
results similar to the present linear response results; the minor
differences may be due in part to the the use of small but finite distortions
in the frozen phonon  calculations to extract the dynamical matrix.
Due to the anharmonic interactions, the observed atomic displacements in
tetragonal structure cannot strictly correspond to the soft phonon TO$_1$
mode in cubic phase. This accounts in part for the difference between
the LDA eigenvectors and the experimental values extracted from the observed
tetragonal structure.

We also performed zone center phonon calculations for the experimental
tetragonal structure.
The phonon frequencies are presented  in
Table~\ref{knbo3-tetra-lto}.
In the tetragonal phase, each  triply degenerate F$_{1u}$ phonon in the cubic
phase
splits into A$_1$+E  modes and the F$_{2u}$ mode  splits into  B$_1$+E modes,
where the doubly degenerate E modes are polarized in \mbox{[1 0 0]} and
\mbox{[0 1 0]}
axes, and the  A$_1$ and B$_1$ modes are polarized along the \mbox{[0 0 1]}
direction.
Compared to the cubic phase, one of the TO$_1$ modes (TO$_1$-A$_1$)
is  stablized due to the tetragonal  distortions, resulting in large
spatial anisotropy  among  the three TO$_1$ components.
Again we used Eq.~(\ref{dm-lo-to}) and the
calculated Born effective charge and dielectric tensors to  obtain  the
frequencies of LO modes.
Since the LO frequencies are no longer isotropic, we
display the calculated values in two different  directions, \mbox{[0 0 1]} and
\mbox{[1 0 0]}, to present the magnitude of this effect.
Except for the soft mode, the calculated frequencies are in good  agreement
with the measurements.

The  linear response calculations for the zone center phonon
modes of the theoretical ground state rhombohedral structure are presented in
Table~\ref{knbo3-r-lto}. The most notable feature is that,  there are no
unstable modes in the rhombohedral phase. In fact none of the modes is
especially soft either. This is consistent with  the observed ground state
structure of KNbO$_3$. It is also consistent with  the disappearance of
two-dimensional X-ray diffuse intensities
in the rhombohedral phase\cite{holma95}.
As in the tetragonal phase,
the LO frequencies depend on the direction
along which they are measured, and
LO frequencies along two directions, \mbox{[1 1 1]} and
\mbox{[ 1 $\bar{1}$ 0]}, are displayed.
The experimental frequencies shown in Table~\ref{knbo3-r-lto}
are from quasi-mode  Raman scattering in the rhombohedral phase\cite{kugel82}.
Nevertheless, they are in reasonable agreement with our calculations.
The eigenvectors of the TO$_1$ modes with frequencies 208 cm$^{-1}$ and  237
cm$^{-1}$
in Table~\ref{knbo3-r-lto} correspond most closely to
the unstable phonon modes in the cubic structure.

\section{Summary}

By minimizing the forces acting on the relaxed atoms,
we have determined the ground state rhombohedral
structure of KNbO$_{3}$,  and the calculated atomic positions
compare well with measurement.
We find that all zone center phonon modes in the rhombohedral structure
are stable and their frequencies  are in good
agreement with experiment.
In the tetragonal phase,
one of the soft zone center modes in the cubic phase
is stablized.
We have calculated the Born effective charge tensors for KNbO$_3$ in
the cubic, experimental tetragonal, and theoretical ground state rhombohedral
structures. The Born effective charges are found to exhibit great sensitivity
to the small atomic displacements in the ferroelectric phase transitions.
The dielectric tensors are also calculated, revealing
about a  \mbox{20$\%$} reduction compared the cubic phase
of the  component in the direction of macroscopic
polarization.

\acknowledgments
Supported by Office of Naval Research grant
N00014-94-1-1044. C.-Z. Wang was supported by National Science
Foundation Grant DMR-9404954. Computations were carried out at the
Cornell Theory Center.

\newpage
 \bibliographystyle{prsty}

\begin{references}
\bibitem{tetra-exp}
  With respect to the reference of Nb atom, the coordinates of the other
  atoms are (from Ref. \cite{hewat73}) (0.5$a$, 0.5$a$, 0.518$c$) for K,
  and (0.5$a$, 0, 0.04$c$),
  (0, 0.5$a$, 0.04$c$) and  (0, 0, 0.544$c$) for O$_1$, O$_2$, and O$_3$,
  respectively. Note that $a$=3.997\AA ~and $c$=4.063\AA ~are lattice
constants.
\end{references}

\begin{table}
  \caption{Ground state structures determined by calculations and
    experiment for  KNbO$_{3}$. The displacements $\delta$
    are in unit of the lattice constant of the  cubic structure.
    The energy difference $(E_{g}-E_{c})$ is in mRy.}
  \vskip 8mm
  \begin{center}
    \begin{tabular}{c|ccc|c}
      $k$-point set & $\delta_{K}$ &  $\delta_{I}$ & $\delta_{II}$ &
$(E_{g}-E_{c})$ \\
      \hline
      (4 4 4) & 0.008 & 0.023 &0.021 & -0.63 \\
      (6 6 6) & 0.009 & 0.026 &0.025 & -1.86 \\
      (8 8 8) & 0.010 & 0.026 & 0.025 & -1.99 \\
      \hline
      LMTO    & $\sim 0.018^{a}$ & $\sim 0.018^{a}$ &
      $\sim 0.018^{a}$ & $\sim$ -1.0$^{a}$ \\
      & $\sim 0.026^{b}$ & $\sim 0.027^{b}$ & $\sim 0.027^{b}$ & $\sim $
-3.0$^{b}$ \\
      \hline
      Expt.$^{c}$ & 0.0130 $\pm$ 81 &0.0333 $\pm$ 15      &0.0301 $\pm$ 9 & \\
    \end{tabular}
  \end{center}
  $^{a}$ Estimated from Postnikov {\it et al.}$^{'}$s calculations at lattice
constant $a$=3.93~\AA~\cite{postni93}.\\
  $^{b}$ Estimated from Postnikov {\it et al.}$^{'}$s calculations at lattice
constant $a$=4.00~\AA~\cite{postni93}.\\
  $^{c}$ Measurement for the rhombohedral phase  at T= 230 K by
Hewat.\cite{hewat73} \\
  \label{ground-s}
\end{table}

\begin{table}
  \caption{Eigenvalues of Born effective charge tensors
    for KNbO$_{3}$ in
    the  cubic, experimental tetragonal, and
    theoretical ground state rhombohedral structures.
    Z$^{*}$(O$_1$) is presented as representative of the three equivalent
oxygen
    atoms in the rhombohedral structure.
    In the experimental tetragonal structure,
    we show only the eigenvalues of Z$^{*}$(O$_1$) and Z$^{*}$(O$_3$),
    since O$_2$ is equivalent to O$_1$ by symmetry. The eigenvectors for
    Z$^{*}$(K)   and Z$^{*}$(Nb) of tetragonal and rhombohedral structures
    are shown as subscripts (see text).}
  \label{z-compare}
  \vskip 6mm
  \begin{center}
    \begin{tabular}{ccccccc}
      &  Present &Present &  Present
      &PW-BP$^{a}$ & PW-BP$^{a}$ & LAPW-BP$^{c}$\\
      \hline
      &    Cubic & Tetragonal & Rhombohedral
      &   Cubic           & Tetragonal$^b$  & Tetragonal \\
      \hline
      Z$^{*}$(K) & 1.12    & 1.12$_{[100]}$ & 1.16$_{[1\bar{1}0]}$
      & 1.14       & ---    & ---  \\
      & 1.12    & 1.12$_{[010]}$ & 1.16$_{[11\bar{2}]}$
      &      1.14     & ---    & ---  \\
      & 1.12    & 1.17$_{[001]}$  & 1.13$_{[111]}$
      &      1.14     & 1.14$_{[001]}$    & 0.82$_{[001]}$   \\
      \hline
      Z$^{*}$(Nb) &9.67 & 9.17$_{[100]}$ & 8.51$_{[1\bar{1}0]}$
      & 9.23          & ---    &  ---    \\
      &9.67 & 9.17$_{[010]}$ & 8.51$_{[11\bar{2}]}$
      & 9.23          & ---    &---\\
      &9.67 & 7.05$_{[001]}$  & 7.47$_{[111]}$
      & 9.23         & 9.36$_{[001]}$    &9.13$_{[001]}$ \\
      \hline
      Z$^{*}$(O)$_{\parallel}$ &  -7.28  & -6.99(O$_{1}$)~~~-5.35(O$_{3}$) &
-6.28(O$_{1}$)
      & -7.01  & -7.10(O$_{3}$)  & -6.58(O$_{3}$)\\
      Z$^{*}$(O)$_{\perp}$     &  -1.74  & -1.77(O$_{1}$)~~~-1.55(O$_{3}$) &
-1.54(O$_{1}$)
      & -1.68  & -1.70(O$_{1}$)  & -1.68(O$_{1}$) \\
      &  -1.74  & -1.40(O$_{1}$)~~~-1.55(O$_{3}$) & -1.55(O$_{1}$)
      & -1.68  & -1.70(O$_{1}$)  & -1.68(O$_{1}$)\\
    \end{tabular}
  \end{center}
  $^{a}$ Plane wave Berry phase calculation by Zhong {\it et
 al.}\cite{zhong-z-94}. \\
  $^b$ Note an ideal tetragonal structure (without observed internal strains)
       was used (private communication with Zhong and Vanderbilt).\\
  $^{c}$ LAPW Berry phase calculation by Resta {\it et al.}\cite{resta93}. \\
\end{table}

\begin{table}
  \caption{Eigenvalues of calculated dielectric tensors $\epsilon_{\infty}$
    of KNbO$_{3}$ in the cubic, experimental tetragonal, and theoretical
    ground state rhombohedral structures.}
  \label{tensor-eps-eig-compare}
  \vskip 6mm
  \begin{center}
    \begin{tabular}{llll}
      cubic     &  6.63    &   6.63 &   6.63   \\
      tetragonal&  6.28$_{[100]}$    &   6.28$_{[010]}$  &  5.07$_{[001]}$  \\
      rhombohedral &5.93$_{[1\bar{1}0]}$    &   5.93$_{[11\bar{2}]}$  &
 5.49$_{[111]}$  \\
    \end{tabular}
  \end{center}
\end{table}

\begin{table}
  \caption{Calculated
    zone center optic phonon frequencies (${\rm cm}^{-1}$) in cubic
    KNbO$_3$.}
  \label{knbo3-c-lto}
  \vskip 6mm
  \begin{center}
    \begin{tabular}{cccccc}
Mode   &   Present & LAPW$^{a}$ & LMTO$^{b}$ & PW$^{c}$         &
Experiment$^{d}$ \\
      \hline
TO$_1$ &    197$i$ & 115$i$ & 203$i$ &143$i$  & soft \\
TO$_2$ &       170 & 168    & 193    & 188    & 198 \\
TO$_3$ &       473 & 483    & 483    & 506    & 521 \\
TO$_4$ &       243 & 266    & 234    &        & 280$^{e}$ \\
      \hline
LO$_1$ &     393$^A$~~~~403$^B$ & & & 407$^B$  & 418 \\
LO$_2$ &     167$^A$~~~~167$^B$ & & & 183$^B$  & 190 \\
LO$_3$ &     757$^A$~~~~888$^B$ & & & 899$^B$  & 826 \\
    \end{tabular}
  \end{center}
  $^{A}$ Obtained with calculated  $\epsilon_{\infty}=6.63$.\\
  $^{B}$ Obtained with  $\epsilon_{\infty}=4.69$,  extracted\cite{zhong-z-94}
from experiment. \\
  $^{a}$ LAPW frozen phonon calculations by Singh {\it et al.}
\cite{singh-b-92}.\\
  $^{b}$ LMTO frozen phonon calculations by Postnikov {\it et al.}
\cite{postni94-1}.\\
  $^{c}$ Planewave frozen phonon calculations by Zhong {\it et al.}
\cite{zhong-z-94}.\\
  $^{d}$ Infrared reflectivity spectroscopy at 710 K by Fontana {\it et al.}
\cite{fontana84}.
 \\ $^{e}$ Measured in the  tetragonal phase, $T = 585$~K.\\
\end{table}

\begin{table}
  \caption{Calculated eigenvectors of the zone center optic  phonon modes
    in cubic KNbO$_3$. O$_{1}$, O$_{2}$ and O$_{3}$  refer to oxygen
    atoms at (0.5, 0, 0)$a$, (0, 0.5, 0)$a$ and (0, 0, 0.5)$a$ respectively.
    Note that we have chosen the representative modes in which the  atoms move
in the
    $\hat{z}$ direction.    }
  \label{knbo3-c-to-vector}
  \vskip 6mm
  \begin{center}
    \begin{tabular}{ccccccc}
        Modes & Frequency(cm$^{-1}$) & K & Nb &       O$_{1}$ & O$_{2}$ &
O$_{3}$ \\
 \hline
       TO$_{1} $ & 197$i$ &  0.01 &  -0.59 &  0.42  &  0.42  &  0.55 \\
       TO$_{2} $ & 170    &  0.88 &  -0.37 & -0.18  & -0.18  & -0.15 \\
       TO$_{3}$  & 473    &  0.02 &  -0.08 &  0.46  &  0.46  & -0.76 \\
       TO$_{4}$  & 243    &  0    &   0    &  1     & -1     &  0   \\
\hline
       LO$_{1}$  & 393    & -0.06 &  -0.38 &  0.63  &  0.63  & -0.23   \\
       LO$_{2}$  & 167    &  0.88 &  -0.45 & -0.11  & -0.11  & -0.09   \\
       LO$_{3}$  & 757    &  0.09 &   0.37 & -0.06  & -0.06  & -0.92  \\
    \end{tabular}
  \end{center}
\end{table}

\begin{table}
  \caption{ Relative atomic displacements in the soft TO$_1$ mode of
    cubic  KNbO$_3$.  Note that the  K atom is taken as the
    reference point and the magnitude of the displacement
    of the Nb atom is taken as the displacement unit.
    The experimental values are extracted from the observed
    tetragonal structure. See text.}
  \label{soft-mode}
  \vskip 6mm
  \begin{center}
    \begin{tabular}{lccccc}
      & K & Nb &       O$_{1}$ & O$_{2}$ & O$_{3}$ \\
      \hline
      Present    &  0  & -1  &   1.6 & 1.6 & 2.2  \\
      LAPW$^a$   &   0  & -1  &   1.4 & 1.4 & 1.9 \\
      PW$^b$     &  0  & -1  &   1.5 & 1.5 & 2.0  \\
      LMTO$^c$   &  0  & -1  &   0.18 & 0.18 & 0.67 \\
      Experiment$^d$ & 0  & -1  &  1.1 & 1.1 & 1.3
    \end{tabular}
  \end{center}
  $^a$ LAPW frozen phonon calculation by Singh and Boyer\cite{singh-b-92}.\\
  $^b$ Plane wave  frozen phonon calculation\cite{zhong-private}. \\
  $^c$ LMTO  frozen phonon calculation\cite{postni94-1}. \\

\end{table}

\begin{table}
\caption  { Calculated zone center phonon frequencies of 
            tetragonal KNbO$_{3}$.  Note that in all A and B modes,
            atoms move in the  $\hat{z}$ direction, $i.e.$ [0 0 1].  }
  \label{knbo3-tetra-lto}
  \vskip 6mm
  \begin{center}
    \begin{tabular}{rccc}
      Mode       &   \multicolumn{3}{c}{Frequency(cm$^{-1}$)} \\
\cline{2-4}
                 &  Present   & Expt$^{a}$ & Expt$^{b}$ \\
      \hline
TO$_1$~~~~~~~E    &  134$i$    & soft         & soft         \\
             A    &306       & 295        & 280         \\
TO$_2$~~~~~~~E    &  174      & 199        &  190         \\
             A    &  171       & 190        &  200             \\
TO$_3$~~~~~~~E    &  479       & 518        & 590         \\
             A    &  613       & 600        & 600         \\
TO$_4$~~~~~~~E    &  253       & 280        & 285         \\
             B    &  262       &            & 290         \\
\hline
LO$_1$~~$[100]$    &  391       & 418        &              \\
         $[001]$   &  401       & 423        & 430          \\
LO$_2$~~$[100]$    &  171       & 191        & 185$^c$         \\
         $[001]$   &  178       & 192        &                 \\
LO$_3$~~$[100]$    &  764       & 822        & 835         \\
         $[001]$   &  834       & 835        & 840           \\
LO$_4$~~$[100]$    &  253       & 279        & 285$^c$       \\
         $[001]$   &  262       &            & 290
    \end{tabular}
  \end{center}
  $^{a}$ Infrared reflectivity spectroscopy at 585 K by Fontana {\it et
al.}\cite{fontana84}\\
  $^{b}$ Roman scattering measurement at 543 K by
          Fontana {\it et al.}\cite{fontana81-pss} \\
  $^{c}$ Measured in [110] direction. \\
\end{table}

\begin{table}
\caption{ Calculated zone center optic phonon frequencies (cm$^{-1}$)
     for the theoretical ground state rhombohedral KNbO$_{3}$ structure.
     Note that the listed experimental frequencies are from quasi-mode Raman
      scattering in the rhombohedral phase.}
  \label{knbo3-r-lto}
  \vskip 6mm
  \begin{center}
    \begin{tabular}{rcc}
  Mode                  & Present &   Experiment$^{a,b}$   \\
\hline
TO$_1$~~~~~~~E~         &      208 &   220           \\
                A$_{1}$ &      237 &   265             \\
TO$_2$~~~~~~~E~         &      170 &   198               \\
               A$_{1}$  &      176 &   198   \\
TO$_3$~~~~~~~E~         &      519 &   536         \\
               A$_{1}$  &      570 &   602           \\
TO$_4$~~~~~~~E~         &      265 &   301     \\
               A$_{2}$  &      238 &   301       \\
\hline
LO$_1$~~$[111]$         &      404 &   423   \\
    $[1\bar{1}0]$       &      389 &         \\
LO$_2$~~$[111]$         &      177 &   198   \\
    $[1\bar{1}0]$       &      172 &         \\
LO$_3$~~$[111]$         &      794 &   837   \\
    $[1\bar{1}0]$       &      803 &         \\
LO$_4$~~$[111]$         &      265 &   301   \\
    $[1\bar{1}0]$       &      265 &         \\
    \end{tabular}
  \end{center}
$^{a}$ Fontana {\it et al.}, Reference\cite{fontana84}.  \\
$^{b}$ Kugel {\it et al.},   Reference\cite{kugel82}.\\
\end{table}

\end{document}